\documentclass{iau_FM}

\newcommand{\hii}{H\,{\sc ii}{}}
\newcommand{\feii}{Fe\,{\sc ii{}}}
\def \zsun{\ifmmode{{\rm\ Z}_\odot}\else{${\rm\ Z}_\odot$}\fi}

\usepackage{graphicx}

\title[SNe~II as metallicity indicators] 
{A correlation between SN~II metal line strength and host \hii\ region oxygen
abundance}

\author[Joseph Anderson]   
{Joseph P. Anderson$^1$,
Claudia P. Guti\'errez$^{1,2,3}$
\and Luc Dessart$^4$}

\affiliation{$^1$European Southern Observatory,
Alonso de Córdova 3107, Casilla 19001,
Santiago, Chile \\email: {janderso@eso.org}\\[\affilskip]
$^2$Millennium Institute of Astrophysics, Casilla 36-D, Santiago, Chile\\[\affilskip]
$^3$Departamento de Astronom\'ia, Universidad de Chile, Camino El Observatorio 1515, Las Condes, Santiago, Chile\\[\affilskip]
$^4$Laboratoire Lagrange, UMR7293, Universite Nice Sophia-Antipolis, 
CNRS, Observatoire de la Cote d'Azur, F-06300 Nice, Franc}

\pubyear{2015}
\jname{Astronomy in Focus, Volume 2} 
\editors{Piero Benvenuti, ed.}
\begin{document}

\maketitle

\begin{abstract}
Dessart et al., demonstrated that type II supernova (SN~II) model spectra present increasing
metal line strength with increasing progenitor metallicity.
To confront these models with observations, we obtained a large sample of
SN~II host \hii\ region emission line spectroscopy.
We show that inferred SN~II host \hii\ region metallicities have a statistically
significant correlation with the strength of SN~II metal lines, specifically \feii\ 5018\AA.
\end{abstract}

Knowledge of the chemical evolution of galaxies is important
for our overall understanding of the Universe.
Outside of our local group of galaxies the dominant source of abundance/metallicity
measurements comes from emission line diagnostics. 
However, there are significant issues with these methods, both
on the absolute abundance scale, and also of large systematic offsets between
different diagnostics (see e.g. \cite[L\'opez-S\'anchez et al. 2012]{lop12}). Hence,
any independent metallicity indicator is of significant value.\\
\indent \cite[Dessart et al. (2014)]{des14} (following the earlier work of \cite[Dessart et al. 2013]{des13})
studied the spectral properties of SN~II ejecta models that resulted from the 
explosion of RSG stars evolved from the main sequence at different metallicities, 
from sub-solar to super-solar.
They found that metal line strengths correlated with 
progenitor metallicity. As an observational follow-up of that work, we obtained host \hii\ region
spectroscopy of a large sample of SNe~II. Using these data, emission line fluxes were extracted and 
host \hii\ region oxygen abundances were determined using the \cite[Marino et al. (2013)]{mar13} O3N2 diagnostic.
For each associated SN~II we measured pseudo equivalent widths (pEWs) of the  \feii\ 5018\AA\ 
absorption line in the SN spectrum interpolated
to 50 days post explosion.\\ 
\indent In Fig.~1a we present a correlation between environment
oxygen abundance (a proxy for metallicity) and SN \feii\ 5018\AA\ pEWs
interpolated to 50 days post explosion. A statistically significant correlation is found in that 
SNe with higher pEWs are found to explode within higher abundance environments.
This initial correlation was achieved with all SNe~II irrespective of their light-curve morphology.
As shown in \cite[Anderson et al. (2014)]{and14} SNe~II show a large range in light-curve morphologies 
(brightness, decline rates, optically thick phase  durations etc). 
Variations in progenitor or explosion properties, 
independent of composition, could produce comparable effects to changes in metallicity.
In Fig.~1b we remove all SNe which we do not consider `normal' SNe~IIP. Within this sub-sample the dispersion is 
significantly reduced. The dispersion
is also reduced when we use a colour in place of time epoch.\\
\indent This work strengthens the conclusions of \cite[Dessart et al. (2014)]{des14}, and motivates
further study to enable SNe~II to be used as independent metallicity indicators throughout the Universe. 
A full analysis will soon be published in Anderson et al., (in preparation).

\textbf{\textit{Acknowledgements:}} we acknowledge the use of `CATS et al.' and Carnegie Supernova Project 
spectroscopy (data to be published in Guti\'erriez et al. in preparation).

\begin{figure}
\centering
\includegraphics[width=6.5cm]{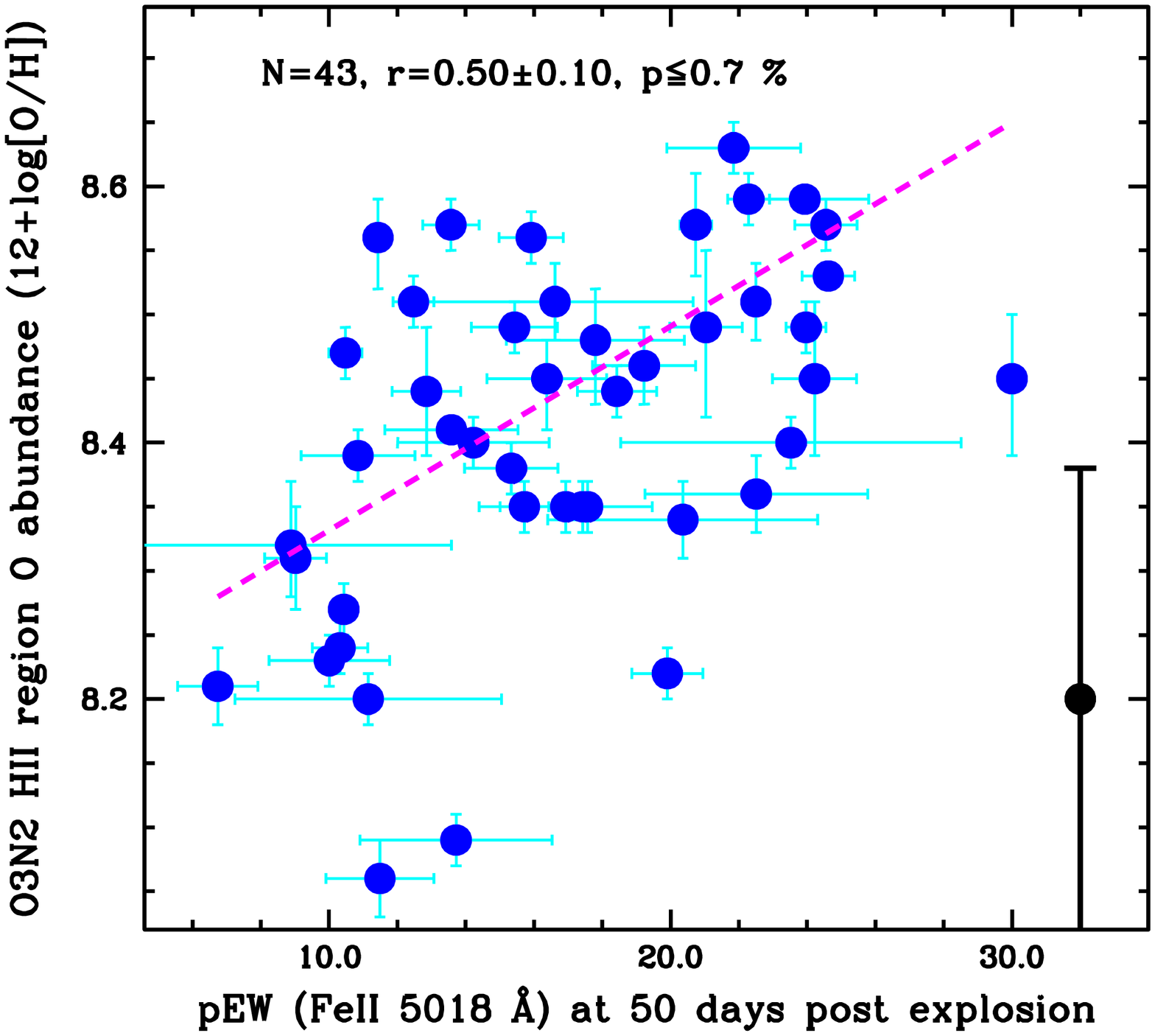} 
\includegraphics[width=6.5cm]{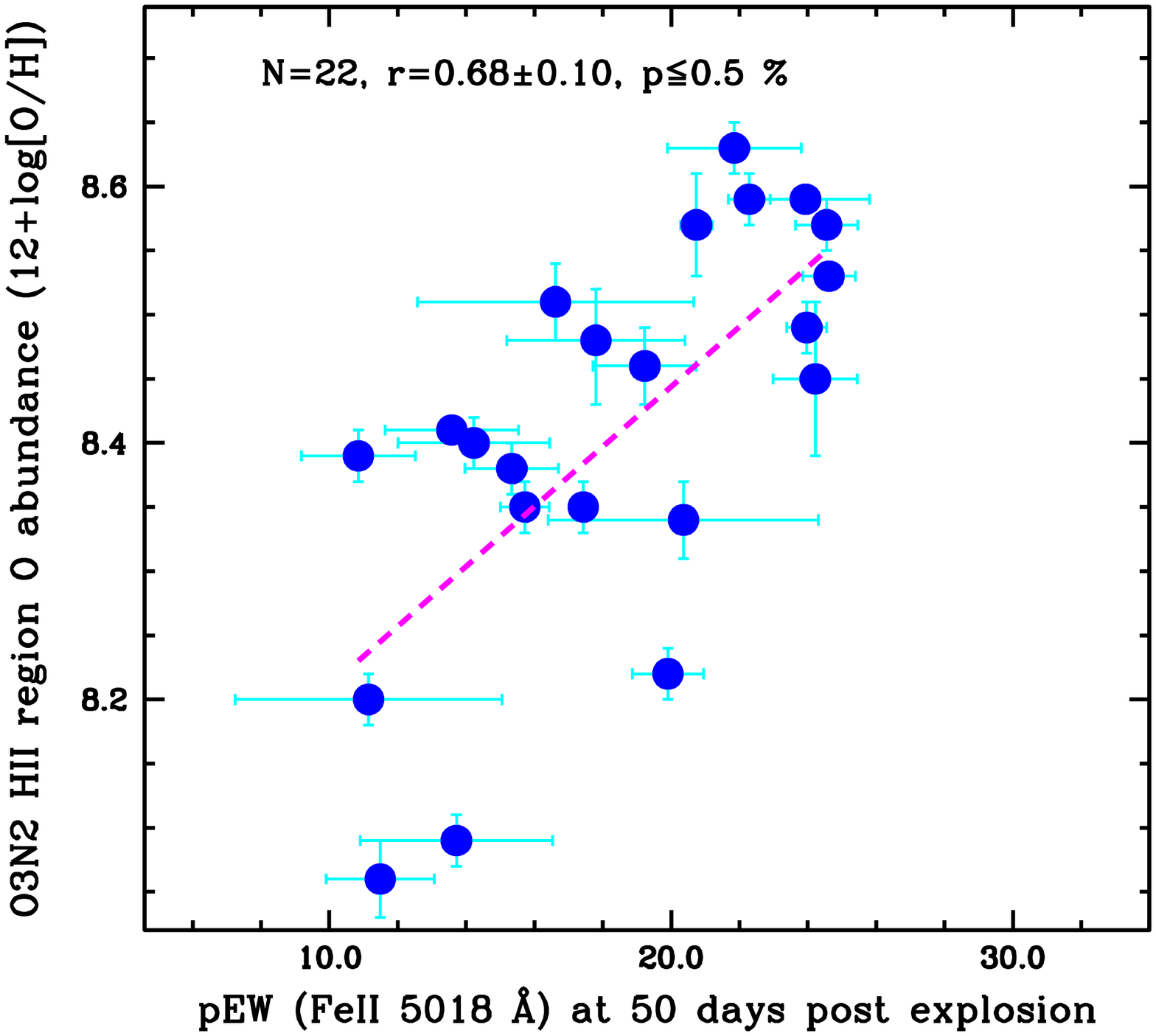} 
\caption{\textit{Left, a):} SN~II pEWs of \feii\ 5018\AA\ at 50 days post explosion
against host \hii\ region oxygen abundance. \textit{Right, b):} same as a) but now for a subset 
of SNe~II whih have light-curve morphologies consistent with `normal' SNe~IIP.}
\end{figure}

\end{document}